\documentclass[iop]{emulateapj}
\pdfoutput=1
\usepackage{graphicx}

\newcommand{\ie}{i.e.,\ }
\newcommand{\eg}{e.g.,\ }

\newcommand{\etal}{et~al.\ }
\newcommand{\ltsima}{$\; \buildrel < \over \sim \;$}
\newcommand{\simlt}{\lower.5ex\hbox{\ltsima}}
\newcommand{\gtsima}{$\; \buildrel > \over \sim \;$}
\newcommand{\simgt}{\lower.5ex\hbox{\gtsima}}
\newcommand{\kms}{km s$^{-1}$}
\newcommand{\magsec}{mag~arcsec$^{-2}$}

\def\muv{$\mu_V$}
\def\mub{$\mu_B$}

\def\bmv{B$-$V}
\def\hi{HI\ }
\def\ha{H$\alpha$\ }
\def\R25{R$_{25}$}
\def\D25{D$_{25}$}
\def\sfr{M$_\sun$~yr$^{-1}$\ }
\def\Msun{M$_\sun$\ }
\def\Lsun{L$_\sun$\ }

\begin{document}

\title{The Extended Optical Disk of M101}

\author{J. Christopher Mihos,\altaffilmark{1} 
	Paul Harding,\altaffilmark{1}
	Chelsea E. Spengler,\altaffilmark{1} \\
	Craig S. Rudick,\altaffilmark{2} and
      John J. Feldmeier\altaffilmark{3}}
\email{mihos@case.edu, paul.harding@case.edu,
        chelsea.spengler@case.edu, craig.rudick@phys.ethz.ch,
        jjfeldmeier@ysu.edu}

\altaffiltext{1}{Department of Astronomy, Case Western Reserve University,
10900 Euclid Ave, Cleveland, OH 44106, USA}

\altaffiltext{2}{Institute of Astronomy, ETH Zurich, CH-8093, Zurich, Switzerland}

\altaffiltext{3}{Department of Physics and Astronomy, Youngstown 
State University, Youngstown, OH 44555, USA}

\begin{abstract}

We have used deep, wide-field optical imaging to study the faint
outskirts of the luminous spiral galaxy M101 (NGC 5457), as well as its
surrounding environment. Over six square degrees, our imaging has a
limiting surface brightness of \mub $\sim$ 29.5 \magsec, and has
revealed the stellar structure of M101's disk out to nearly 25\arcmin\ 
(50 kpc), three times our measured \R25 isophotal size of the optical
disk. At these radii, the well-known asymmetry of the inner disk slews
180 degrees, resulting in an asymmetric plume of light at large radius
which follows the very extended HI disk to the northeast of M101. This
plume has very blue colors (\bmv $\sim$0.2), suggesting it is the
somewhat more evolved (few hundred Myr to $\sim$1 Gyr) counterpart of
the young far ultraviolet emitting population traced by GALEX imaging.
We also detect another, redder spur of extended light to the east of the
disk, and both structures are reminiscent of features produced during
fly-by galaxy interactions. However, we see no evidence of very extended
tidal tails around M101 or any of its companions which might be expected
from a recent encounter with a massive companion. We consider the
properties of M101's outer disk in light of possible past interactions
with the nearby companion galaxies NGC 5477 and NGC 5474. The detection
of optical starlight at such large radii gives us the ability to study
star formation histories and stellar populations in outer disks over a
longer timescales than those traced by the UV or \ha emitting
populations. Our data suggest ongoing buildup of the M101's outer disk
due to encounters in the group environment triggering extended star
formation and tidal heating of existing disk populations.

\end{abstract}


\section{Introduction}

The diffuse, low surface brightness outskirts of galaxies hold a
remarkable range of information about processes driving galaxy evolution.
As galaxies are thought to grow ``inside-out," where the central regions
form first, and the outskirts later, the most recent signatures of
galaxy assembly should lie in their faint outer reaches.
Interactions and accretion leave behind long-lived tidal tails and
stellar streams (\eg Arp 1966; Toomre \& Toomre 1972; Bullock \&
Johnston 2005; Mart\'inez-Delgado \etal 2010). Star formation in the
outer disks of galaxies probes the mechanisms for star formation at low
gas density (Kennicutt \etal 1989; Martin \& Kennicutt 2001; Bigiel \etal
2008). Dynamical models show that substantial radial migration of stars
can occur, whereby stars formed in the inner disk can move outwards and
populate the disk outskirts (Sellwood \& Binney 2002; Debattista \etal
2006; Roskar \etal 2008ab). All of these different processes leave
signatures in the structure, stellar populations, and kinematics of the
outer disk which can be used to develop a more complete picture of disk
galaxy evolution.

The nearby spiral galaxy M101 (NGC 5457) presents an opportunity to
study the outer disk of a giant Sc galaxy in detail. M101 is the
dominant member of a small group of galaxies (Geller \& Huchra 1983;
Tully 1988), and as its well-known asymmetry attests to recent or on-going
interactions with its companions (Beale \& Davies 1969; Rownd \etal
1994; Waller \etal 1997). As the small group environment is the most
common for galaxies, the processes shaping M101 have likely shaped the
bulk of the galaxy population as well. Furthermore, M101 is well-studied
at many wavelengths, giving us a comprehensive view of the structure and
kinematics of its baryonic components. It is also close enough that deep
space-based imaging has the potential to image its stellar populations
directly, raising the possibility of obtaining direct information about
the distribution of stellar age and metallicity in its disk (see, \eg
similar studies for other nearby galaxies by de Mello \etal 2008;
Dalcanton \etal 2009, 2012; Radburn-Smith \etal 2011, 2012).

A variety of studies have shown that M101's disk extends well beyond its
canonical \R25 optical radius. Deep 21-cm mapping has revealed an
extended, asymmetric gaseous disk (van der Hulst \& Sancisi 1988; Walter
\etal 2008), with a plume of extended HI at very low column density
traced out to nearly 100 kpc in extent (Huchtmeier \& Witzel 1979; Mihos
\etal 2012). GALEX imaging has shown that far-ultraviolet (FUV) emission
extends to very large radius (Thilker \etal 2007), indicating that star
formation is taking place in the outer disk, well beyond the radius at
which star formation is thought to be suppressed due to low gas densities
(\eg Kennicutt 1989). Whether this star formation has only been recently
triggered, perhaps by an interaction, or reflects an ongoing process of
disk building requires a more detailed understanding of the underlying
stellar populations in the outer disk.

While this extended star formation adds young stellar populations to the
outer disk, interactions with companion galaxies will distort and
tidally heat the inner disk, growing the outskirts of the disk by moving
inner disk stars outwards. Tidal stripping of the interacting companions
can directly deposit stars in the outer disk as well. Over longer
timescales, M101's strong asymmetry and marked spiral structure make it
a conducive environment for radial migration processes, which also drive
old stars outwards. All these processes can lead to changes in the
surface brightness and color profiles of the disk, and may be
responsible for the diverse surface brightness profiles seen in disk
galaxies (van der Kruit 1979; Pohlen \etal 2002, 2007). While late-type
spirals such as M101 are typically associated with down-bending or
truncated surface brightness profiles (Pohlen \& Trujillo 2006), these
processes may build the outer disk and transform their profiles into the
up-bending or anti-truncated profiles characteristic of early type
disks. If dynamical processes in the group environment also lead to
changes in the Hubble type of the galaxy (Tran \etal 2001; McGee \etal
2008; Kova\v c \etal 2010), this could present a self-consistent
explanation for the correlation between disk profile and Hubble type
seen in studies of disk galaxies (Pohlen \& Trujillo 2006; Erwin \etal
2008). In M101, we have an opportunity to examine a
galaxy in the throes of environmentally-driven evolution, and to study
disk building and the response of the luminosity profile and stellar
populations in its outer disk.

Interaction signatures abound in both M101's morphology and kinematics.
The strong asymmetry of the galaxy suggests a recent interaction with at
least a moderately massive galaxy (Beale \& Davies 1969; Waller \etal
1997), while high velocity gas complexes in the HI distribution are
thought to be either signatures of gaseous accretion onto the disk (van
der Hulst \& Sancisi 1988) or induced by a recent interaction (Combes
1991). A number of nearby companions exist as possible interaction
partners, including NGC 5477 and NGC 5474, but a ``smoking gun''
signature of their interaction with M101 remains missing.  Both the
HI plume to the southwest of M101 and the diffuse HI found between 
M101 and NGC 5474 (Huchtmeier \& Witzel 1979; Mihos
\etal 2012) are suggestive of tidal debris, although the low spatial
resolution of the 21-cm data ($\sim$ 9\arcmin) make that determination
difficult. Deep optical imaging of other nearby disk galaxies has
revealed many examples of extended low surface brightness tidal streams
which trace an interaction (Mart\'inez-Delgado \etal 2010), motivating a
deeper search for optical tidal features around M101 and its companions.

However, while the information content in the outskirts of disk galaxies
is large, accessing it proves difficult, particularly for nearby
galaxies such as M101\footnote{While the exact distance to M101 remains
uncertain (see Matheson \etal 2012 and references therein), in our
analysis we adopt a distance of 6.9 Mpc; at this distance, 1\arcmin =
2.0 kpc.}. The surface brightness of the outer disk is 10-100 times
fainter than the night sky background, and at the distance of M101 the
disk and any associated tidal features could potentially subtend a half
a degree or more in size. Furthermore, the long dynamical timescales in
the outer disk ($\sim$ 0.5 Gyr) mean that photometric structures are
likely not well-mixed azimuthally, so that the standard practice of
azimuthally averaging the light to build signal to noise has the
potential to ``average away'' unmixed substructure and yield misleading
results. To study these regions in detail demands deep, wide-field
imaging sensitive enough to produce two dimensional maps of luminosity
and color without the need to azimuthally average.

Here we report deep ($\mu_{B,lim}=29.5$), wide-field (6 square degree)
imaging of M101 and its surrounding environment. We use this dataset to
search for interaction signatures around M101 and its companions and to
study the structure and color of M101's disk at large radius. Coupled
with extant multiwavelength datasets, our deep optical imaging places
constraints on the star formation history and mix of stellar populations
in the outer disk, and probes mechanisms for galaxy disk building in the
group environment.

\section{Observations}

We observed M101 with Case Western Reserve University's Burrell Schmidt
telescope in April 2009 and April 2010. Full details of our
observational technique can be found in Mihos \etal (2005), Rudick \etal
(2010), and Mihos \etal in preparation; we emphasize the most important
points here, as well as any updated techniques adopted. The Burrell
Schmidt images a 1.5\degr x1.5\degr\ field of view onto a 4Kx4K SITe
CCD, yielding a pixel scale of 1.45\arcsec\ pixel$^{-1}$. 
The April 2010 data were taken in Washington M, similar to Johnson
V, but $\sim$ 300\AA\ bluer, which cuts out the \ion{O}{1} $\lambda$5577
night sky emission line and reduces variability in the sky levels (see
Feldmeier \etal 2002). To give an adequate spectral baseline to measure
meaningful colors, the April 2009 data were taken through a custom-designed
filter similar to Johnson B, but with a central wavelength $\sim$ 200\AA\
bluer.  In both cases the data were taken during moonless
photometric nights where the frame-to-frame variation in the photometric
zeropoints agreed to within a 1$\sigma$ scatter of 0.01 magnitude after
a constant extinction correction with airmass was applied. A total of 61
and 60 images of M101 were taken in B and M, respectively, with the
pointings dithered by up to 0.5\degr\ to minimize uncertainties due to
large scale flat fielding effects. We also observed offset blank sky
pointings (40 in B and 55 in M), bracketing the M101 exposures in time
and hour angle, to construct dark sky flats for the dataset. In B, both
the M101 and sky pointings were 1200s exposures, yielding 500--700
counts pixel$^{-1}$ in the sky. In M, the exposures were 900s, resulting
in sky levels of 1200--1400 counts pixel$^{-1}$.

\begin{deluxetable}{lccc}
\tabletypesize{\scriptsize}
\tablewidth{0pt}
\tablecaption{Properties of M101 and Nearby Companions}
\tablehead{\colhead{ } & \colhead{M101} & \colhead{N5477} & \colhead{N5474}}
\startdata
RA & 14:03:12.5 & 14:05:33.3 & 14:05:01.6 \\
Dec & +54:20:56 & +54:27:40 & +53:39:44 \\
Type & SAB(rs)cd & SA(s)m & SA(s)cd pec \\
$V_{\rm helio}$ [\kms] & 241 & 304 & 273 \\
Distance [Mpc]                 & 6.9 & ... & ... \\
R$_{\rm proj}$ [arcmin] & ... & 22 & 44 \\
R$_{\rm proj}$ [kpc] & ... & 44 & 88 \\
$M_B$              & $-21.0$ & -15.0 & -17.9 \\
$L_B$  [$10^{10}$ \Lsun]             & 3.9 & 0.015 & 0.22 \\
$B-V$              & 0.44 & 0.35 & 0.48 \\
\R25 (RC3) [arcmin] & 14.4 & 0.8 & 2.4 \\
\R25 (this study) [arcmin] & 8.0 & 0.85 & 2.1 \\
$V_c$ [\kms] & 250 & ... & 40 \\
$M_{HI}$  [$10^{9}$ \Msun]        & 24.1 & 0.18 & 1.50
\enddata
\tablecomments{Hubble type, photometric properties, and sizes of the
galaxies taken from the RC3 (de Vaucouleurs \etal 1991). Rotation speeds
taken from Zaritsky \etal 1990 (M101) and Rownd \etal 1994 (N5474). HI masses
taken from Huchtmeier \& Richter (1988). The physical distance to M101
is adopted from Matheson \etal 2012 and references therein. R$_{\rm
proj}$ is the projected distance from M101.}
\label{regprop}
\end{deluxetable}

\begin{figure*}[]
\centerline{\includegraphics[width=6.5truein]{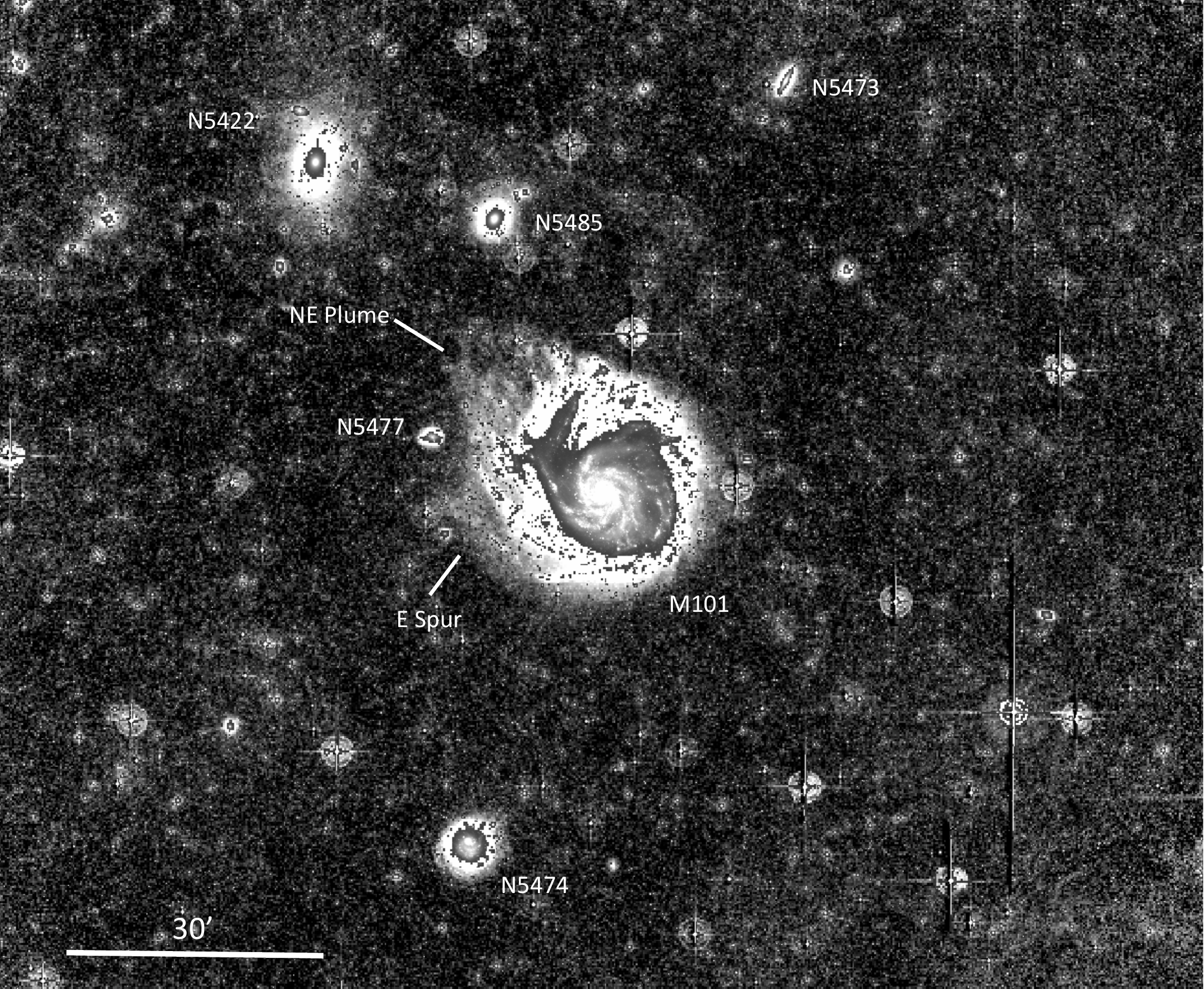}}
\caption{Final B mosaic, spanning 2.5\degr\ (300 kpc) across, with a limiting
surface brightness of \mub=29.5 \magsec. North is up
and east is to the left. In this image, regions of high surface brightness
(\mub$<25.5$) are shown at native resolution (1.45\arcsec/pixel) and
rescaled in intensity to show the bright inner regions of galaxies. At
lower surface brightnesses, the image has been masked and median binned
into 9x9 pixel (13\arcsec x13\arcsec) bins. Bright galaxies in the field are
labeled, as are the NE Plume and E Spur, two features in the outer disk which are
discussed in detail in Section 3.3.}
\label{Bmosaic}
\end{figure*}

The offset sky pointings were used to construct flat fields for each
season's dataset, using the techniques described in Rudick \etal (2010).
In short, each sky pointing is masked of stars and objects using
IRAF's\footnote{IRAF is distributed by the National Optical Astronomy
Observatory, which is operated by the Association of Universities for
Research in Astronomy (AURA) under cooperative agreement with the
National Science Foundation.} {\tt objmask} task, then binned into 32
pixel blocks so that any extended low surface brightness contaminants
(reflections, extended wings of bright stars, etc) could be identified
and masked manually. After masking, the sky pointings are median
combined to produce a preliminary flat; this flat is then used to
flatten the individual skies. Once flattened, we fit and remove a sky
plane from each sky, and then reconstruct an updated flat field. We
repeat this process five times, at which point the flat field converges;
the final flat field is then used to flatten the individual M101 images.

After flat fielding, we subtract the wings of bright stars using a PSF
model which included positionally dependent term to account for internal
reflections (see Slater \etal 2009 for details). For sky subtraction, we
note that even as large as M101 itself is, our very large field of view
gives plenty of sky area around the outskirts to measure the sky level.
We measure this sky level by first masking M101 and the surrounding
bright galaxies, and using IRAF's objmask task to mask fainter stars and
background galaxies in the field. We then rebin the images by
calculating the median in 32x32 pixel bins (excluding masked pixels from
the median), and fit a sky plane to each rebinned image, which is
subsequently subtracted from the original image. Finally, all images are
registered, scaled to a common photometric zeropoint, and medianed
together to form the final image. The photometric zeropoint for each
frame is determined from SDSS photometry of stars, using the photometric
conversion of Lupton (2005) to convert from {\it ugriz} to Johnson B and
V. Comparing our photometry to transformed B,V photometry for SDSS stars
in the field, we recover magnitudes to with $\sigma_V = 0.025$
magnitudes and B-V colors to $\sigma_{B-V}=0.04$ magnitudes for SDSS
stars in the magnitude range V=14.5--17 and color range \bmv=0.0--1.5.
 
After the master image is constructed, we also create a binned version
of the image to maximize signal to noise at low surface brightness. To
make this image, we first use IRAF's {\tt objmasks} task to mask bright
objects in the field -- stars, bright galaxies, and background sources.
We then rebin the image into 9x9 pixel (13\arcsec x13\arcsec or 450x450
pc) blocks, calculating the median intensity of unmasked pixels in each
block. If more than 50\% of the pixels in a given block are masked, we
leave the entire block masked so that a small number of unmasked pixels
won't weigh disproportionately large in the final map.

In the analysis that follows, we calculate surface brightnesses using 
``asinh magnitudes'' (Lupton \etal 1999):
\begin{eqnarray}
\nonumber
\mu_B &=& ZP_B - 2.5\log(b_B) - a \sinh^{-1}(f_B/2b_B) - C_B(B-V) \\
\nonumber
\mu_V &=& ZP_V - 2.5\log(b_V) - a \sinh^{-1}(f_V/2b_V) - C_V(B-V)
\end{eqnarray}
where $a=1.08574$, and for each image, $f$ is counts in ADU, $ZP$ is the
photometric zeropoint, $b$ is the noise in the image, and $C$ is the
filter color term. At high flux levels, asinh magnitudes are equivalent
to regular magnitudes, while at the faint end, near the noise level in
the images, asinh magnitudes are better behaved as the flux levels drop
through zero. The photometric zeropoints ($ZP_B=29.73, ZP_V=29.62$) and color
terms ($C_B=0.121, C_V=0.265$) are derived from the final mosaic using
SDSS stars in the field and the Lupton (2005) conversion from {\it ugriz} to B
and V. Since at faint light levels we will be measuring surface
brightness from the 9x9 binned image, the image noise parameter $b$ is
determined by the noise in this rebinned image. We measure these values
by first creating histograms of the bin intensities in sky regions away
from bright galaxies, and then fit Gaussians to the negative intensity
side of the histograms. We derive values of 0.7 and 0.8 ADU for $b_B$
and $b_V$, respectively. Given these measures, our 1$\sigma$ limiting
surface brightnesses are $\mu_B=29.5, \mu_V=29.0$. During the analysis
which follows, we correct the photometry for galactic extinction using
values of $A_B=0.031, A_V=0.023$ (Schlafly \& Finkbeiner 2011).

\section{Results}

\subsection{Morphology and multiwavelength comparison}

Figure \ref{Bmosaic} shows our final B mosaic, spanning $\sim$ 2.5\degr\
(300 kpc) on a side. In the image shown in Figure \ref{Bmosaic}, we have
replaced the masked inner regions of galaxies with the original pixels,
rescaled in intensity to show their high surface brightness (\mub $<$
25.5 \magsec) structure. The three galaxies north of M101 --- NGC 5422
(1820 \kms), NGC 5485 (2000 \kms) and NGC 5473 (2026 \kms) --- are
background objects, while NGC 5477 (304 \kms) 22\arcmin\ (44 kpc) to the
west and NGC 5474 (273 \kms) 44\arcmin\ (88 kpc) to the south are likely
physically associated with M101 (241 \kms). Table 1 summarizes the basic
properties of M101 and its companions NGC 5474 and NGC 5477.

The well-known inner asymmetry of M101's disk at high surface brightness
is easily seen as an extension to the southwest in the rescaled high
surface brightness parts of the image. Interestingly, the radius of the
\muv=26.5 isophote (approximately where the colorbar saturates to white
in Figure \ref{Bmosaic}) is actually quite circular, with a radius of
12.5\arcmin, but centered $\sim$1.6\arcmin\ north of M101's nucleus. At
larger radius and even lower surface brightness, the asymmetry slews
180\degr\ from the inner asymmetry, as M101's disk displays an
asymmetric plume of light to the northeast. In this plume, which we
refer to hereafter as the NE Plume, we can trace optical starlight out
to 24\arcmin\ (48 kpc) from the center of M101. We also find a spur of
optical light extending 19\arcmin\ (38 kpc) to the east of M101,
referred to in this paper as the E Spur. The dwarf companion NGC 5477
sits between these two features, in a gap reminiscent of those seen in
simulations of disk galaxies interacting with dwarf companions (\eg
Hernquist \& Mihos 1995; Walker \etal 1996).

Aside from these features, down to our limiting surface brightness of
\mub=29.5, we see no evidence of any long, extended tidal tails around
M101 which might signal a strong prograde interaction in its recent
history. To the south, NGC 5474 is often thought to be the interacting
companion driving M101's asymmetry, due to its own asymmetric nature
(\eg Kornreich \etal 1998) and the presence of diffuse HI
between the galaxies (Huchtmeier \& Witzel 1979; Mihos \etal 2012).
However, NGC 5474's outer isophotes are remarkably round, with no
distortion at large radius indicating any strong tidal interaction. This
lack of irregular isophotes at large radius is particularly noteworthy
given the significant offset (1.25 kpc, or 0.3\R25) of NGC 5474's bulge
from disk of the galaxy (van der Hulst \& Huchtmeier 1979; Kornreich
\etal 1998). Because the lifetime of an offset nucleus should be short,
on the order of a few dynamical times ($\sim$ a few $\times 10^8$ Myr),
this argues that any interaction driving this morphology is ongoing or
recent. In such a case, one would expect the tidal interaction
signatures to be even stronger in the outskirts of the galaxy, where
material is more easily perturbed during an interaction. Furthermore,
the HI kinematics of the galaxy show simple rotation, with no sign that
the bulge plays any dynamical role (van der Hulst \& Huchtmeier 1979;
Rownd \etal 1994). While the bulge could be a companion galaxy merely
projected onto the disk of NGC 5474, the fact that the spiral arms
appear symmetric around the bulge suggest that the disk and bulge are
dynamically linked. How such a system could develop with no obvious
signs of optical irregularity at large radius remains a mystery.

\begin{figure*}[]
\centerline{\includegraphics[width=6.0truein]{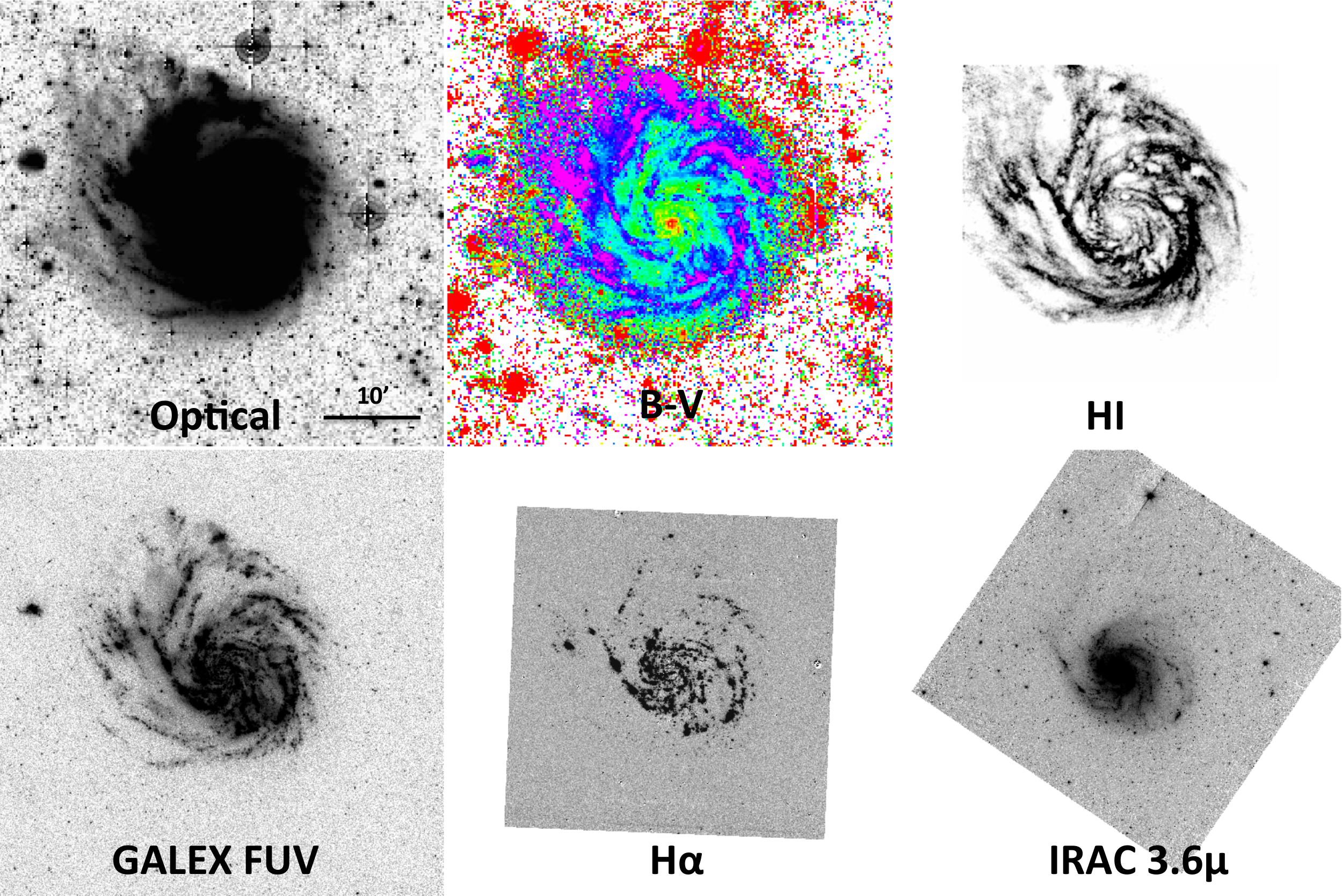}}
\caption{Multiwavelength view of M101. Top left: Our B-band image. Top
middle: Our B-V color map, scaled from \bmv$<$0.2 (magenta) to
\bmv$>$0.7) (red). Top right: THINGS neutral hydrogen map (Walter \etal
2008). Bottom left: GALEX near ultraviolet map (Gil de Paz \etal (2007).
Bottom middle: H$\alpha$ map (Hoopes \etal 2001). Bottom right: IRAC
3.6$\micron$ map (Dale \etal 2009). All images have been registered to
the same orientation and spatial scale as the optical image, with north
up and east to the left.}
\label{M101multi}
\end{figure*}

We present a multiwavelength montage of M101 in Figure \ref{M101multi}.
The scaling of the B-band image is meant to show the faint features in
the outer disk, and saturates to black at surface brightnesses brighter
than \mub=26. The \bmv\ color map is constructed from our deep B and V
images, and shows colors down to a magnitude limit of \muv = 28.75. In
the inner disk, we see a color gradient (discussed quantitatively in Section 3.2)
from the red nucleus to the bluer outskirts of the disk. The NE Plume is
clearly quite blue, with many pixels with \bmv\ colors $<$ 0.3, even in
low surface brightness regions well away from the bright spiral arms. We
emphasize that each of the binned pixels in our optical images are
binned {\it independently}, so that the spatial clumping of these blue
pixels shows the robustness of the detection, rather than being an
artifact of any spatial smoothing. Regions in the E Spur have a redder
B-V color of $\sim$ 0.4--0.5, likely reflecting an older stellar
population in that region. We quantify the photometric properties of the
E Spur and NE Plume in more detail in Section 3.3.

Comparing our optical imaging to the THINGS \hi map (Walter \etal 2008),
it is clear that we detect starlight as far out as the \hi is detected
--- nearly 50 kpc from the center of M101. The gross morphology of M101
in neutral hydrogen and starlight is strikingly similar: the \hi
features trace the optical starlight in both the NE Plume and the E
Spur, and the sharp decline in \hi surface density seen to the SW is
matched by a very sharp decline in the optical light in those regions.
Co-spatial with both the NE Plume and E Spur, high velocity gas is
observed in \hi (van der Hulst \& Sancisi 1988; Kamphuis 2008; Walter
\etal 2008; Mihos \etal 2012), arguing that these features arise from
some strong dynamical forcing with a component out of the plane of
M101's disk. Beyond the coverage of the THINGS map, deep single-dish
imaging has also revealed in a long ($\sim$ 48\arcmin\ or 96 kpc) plume
of \hi\ to the SW at very low column density as well (Huchtmeier \&
Witzel 1979; van der Hulst \& Sancisi 1988; Mihos \etal 2012). However,
in the wide-field image of M101 presented in Figure \ref{Bmosaic}, we
see no sign of any broadband starlight associated with this extended \hi
plume.

The GALEX far ultraviolet map (Gil de Paz \etal 2007) shown in Figure
\ref{M101multi} traces emission from hot stars with ages up to $\sim$
100 Myr (\eg Leitherer \etal 1999). M101 is classified as a Type I XUV
disk (Thilker \etal 2007), displaying extended, structured UV emission
well beyond the nominal optical radius. This extended FUV emission
roughly follows the optical light down to surface brightnesses of \mub
$\sim$ 27, but at fainter surface brightnesses we trace starlight even
further out, in regions with little sign of FUV emission. The \ha map
(Hoopes \etal 2001) traces even younger stars, with ages $\sim$ 10 Myr,
effectively giving a measure of the instantaneous star formation rate.
Again, we find very little \ha emission in the very extended features of
M101's disk --- the E Spur and NE Plume each host only two small
\ion{H}{2} regions within their area. Other than these \ion{H}{2}
regions, we see no \ha emission in the extended disk, down to the
sensitivity limit of $1.6\times10^{-17}$ erg s$^{-1}$ cm$^{-2}$
arcsec$^{-2}$ (or $6\times10^{-4}$ \sfr\ kpc$^{-2}$ using the
L(\ha\llap) to SFR conversion of Hunter \etal 2010).

Finally, the Spitzer IRAC 3.6$\micron$ map (Dale \etal 2009) can also be
seen in Figure \ref{M101multi}. Unfortunately, the outermost regions
of the optical disk extend off the IRAC pointing and the imaging does
not go very deep --- it has a limiting surface brightness of $\sim$ 0.004
MJy sr$^{-1}$, at least an order of magnitude brighter than needed to
detect the low surface brightness features at large radius (see \eg the
discussion in Krick \etal 2011). We include it for completeness of
multiwavelength montage presented in Figure~\ref{M101multi}, but do not
use it in any of our subsequent analysis.

\subsection{Quantitative Structure and Photometric Analysis}

Analyzing the structure of M101's disk over such a broad range of
surface brightness and radial extent takes some care. In the faint outer
parts of the disk, background objects and faint foreground stars can
dominate the light from M101's disk, and need to be masked before
analysis. Furthermore, to maximize signal to noise, we want to median
smooth the image after masking. However, in the inner portions of the
disk, masking bright sources can eliminate star forming regions of the
disk and systematically bias the profile to lower surface brightnesses
and redder colors. We therefore take a hybrid approach, where our
analyses in the inner disk use the full resolution, unmasked images,
while for the outer disk we run the analysis on the images after being
masked and rebinned. We choose the transition radius to be 8\arcmin,
which corresponds to surface brightnesses of roughly \muv $\sim$ 25, and
find no sharp discontinuity in the extracted profiles at this radius due
to the change in technique.

\begin{figure*}[]
\centerline{\includegraphics[width=6.0truein]{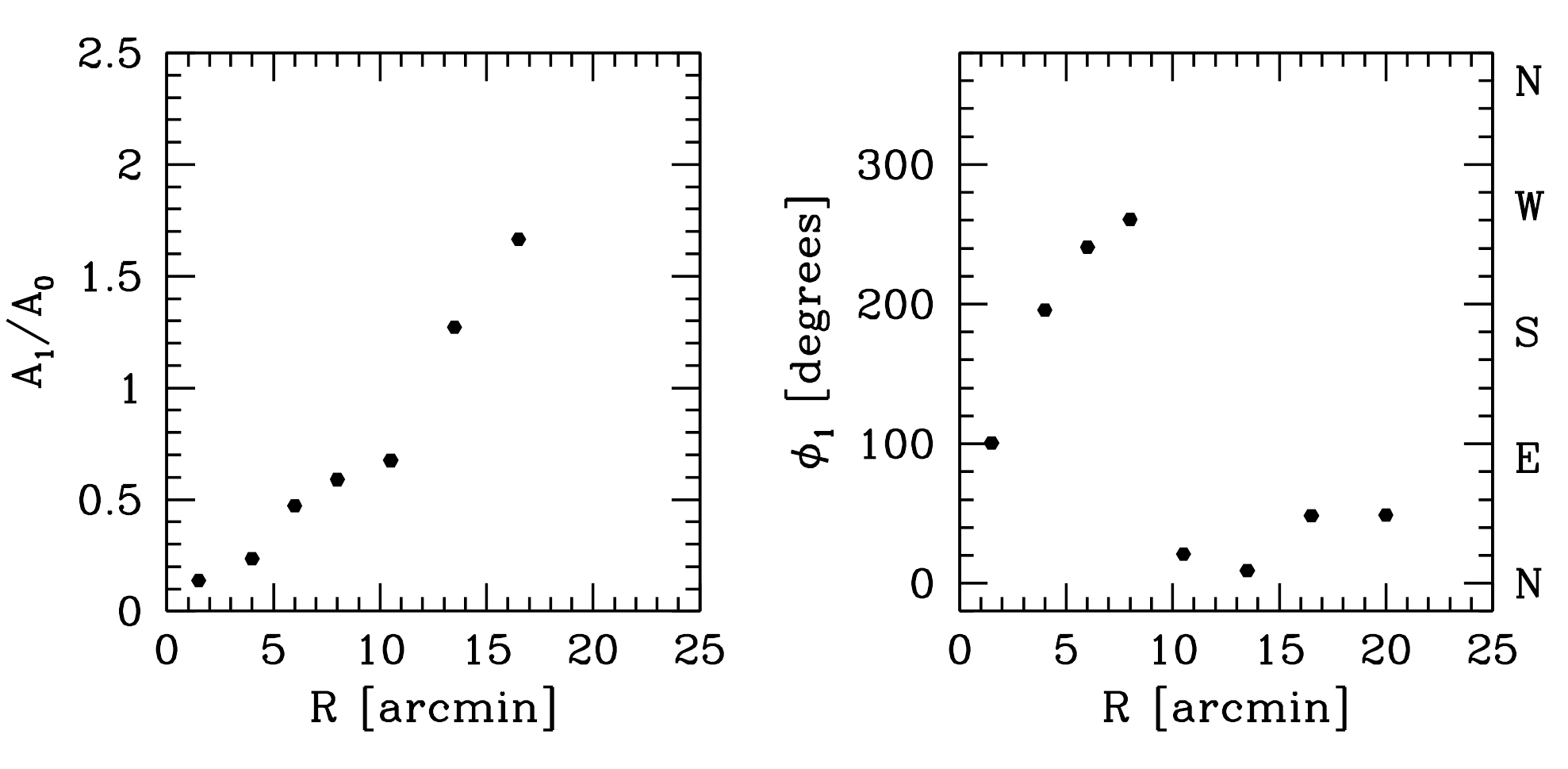}}
\caption{Left: The strength of the $m=1$ asymmetry (relative to $m=0$)
in M101's disk, as a function of radius. Right: The position angle of
the $m=1$ maximum (measured north through east), as a function of
radius.}
\label{mofr}
\end{figure*}

We start our analysis by quantifying the strength and orientation of the
$m=1$ lopsided mode in the disk. As a function of radius, we decompose
the azimuthal surface brightness into Fourier modes, \ie $$ I_B(\theta)
= \sum_m A_m \cos(m\theta + \phi_m)$$ As with the surface brightness
profile, the analysis is done on the unbinned, unmasked map at
R$<8$\arcmin, and on the binned, masked map at larger radii. The
amplitude and position angle of the $m=1$ mode is shown in Figure
\ref{mofr}. The amplitude increases continually with radius, such that
outside R=12\arcmin\ the disk is dominated by the $m=1$ mode. The
position angle of the lopsided mode swings from the SW ($\phi_1$ =
180\degr--270\degr) in the inner high surface brightness regions to a
NNE position ($\phi_1 \sim$ 10\degr--50\degr) in the outskirts, where
the NE Plume dominates the structure of the disk.

\begin{figure*}[]
\centerline{\includegraphics[width=5.0truein]{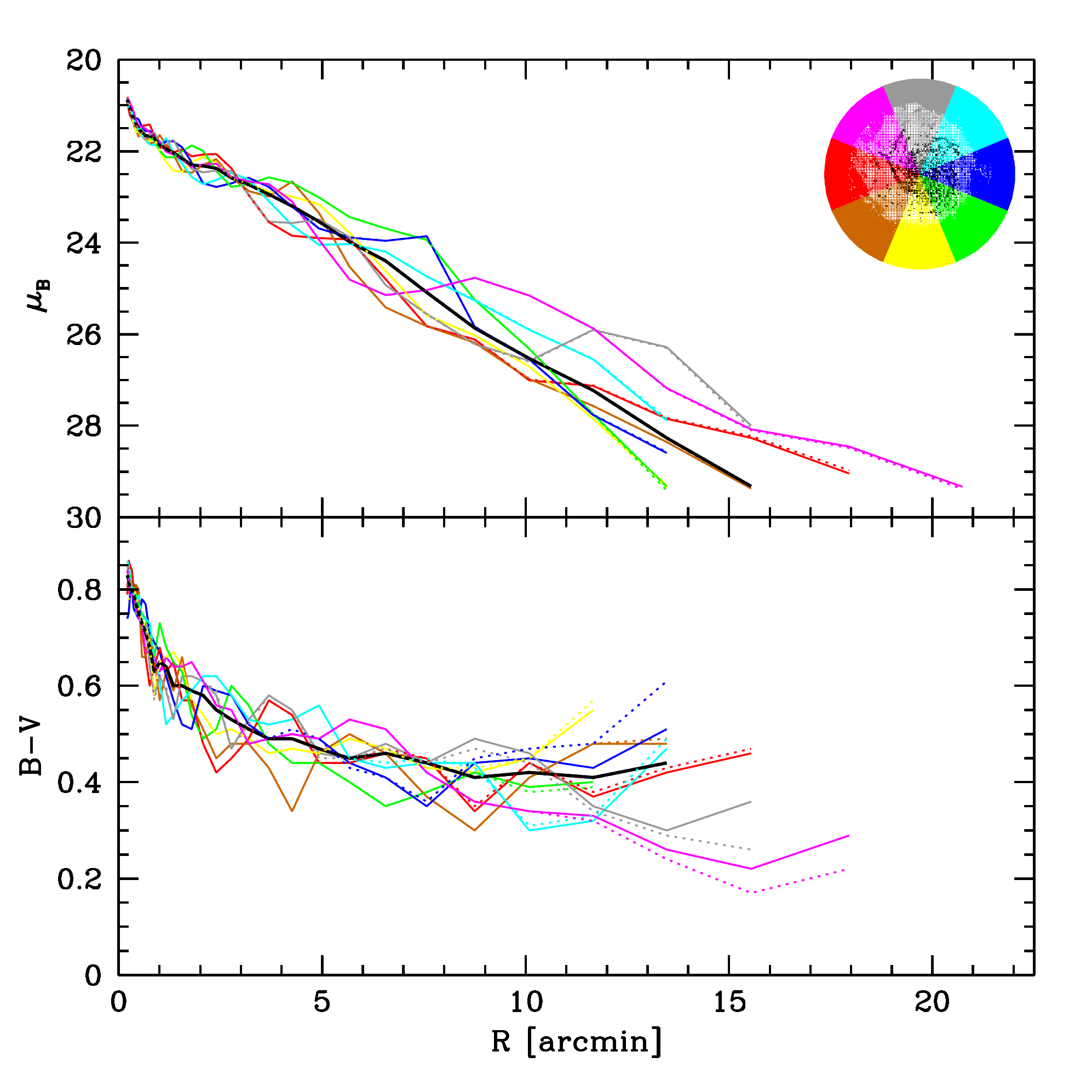}}
\caption{Radial profiles of M101 taken along angular wedges spanning
45\degr\ wide. Top panel: B-band surface brightness profile. Bottom
panel: \bmv color profile. The color coding of the lines matches the
schematic picture of M101 shown in the upper panel, with north being up
and east to the left. The heavy black line in each panel indicates the azimuthally averaged
profile. The major source of uncertainty in the profiles is due to sky subtraction;
to illustrate this effect, the profiles are computed using two measurements
of the background correction, a global correction (solid curves) and a
local correction (dashed curves; see the text for details). The differences in
the profiles calculated in these two ways are noticeable only in the outermost
radial bins at low surface brightness.
}
\label{radprof}
\end{figure*}

Because of the strong asymmetries in the disk of M101, rather than
conduct azimuthally averaged surface photometry, we extract surface
brightness and color profiles in eight octants around the disk, as well
as constructing an azimuthally averaged profile, all shown in Figure \ref{radprof}.
At low surface brightness, background estimation and subtraction becomes
increasingly important and, on large angular scales, harder to quantify
due to uncertainty in sky subtraction as well as structure in the
background. We estimate the background correction in two ways: a global
background defined as the median pixel intensity in an annulus from
R=25\arcmin--42\arcmin\ around M101, and, for each octant, a local
background defined by the median pixel intensity in the associated
octant of the background annulus. We then calculate surface brightnesses
profiles using after correcting for each background measure and plot
both measures in Figure \ref{radprof}; the solid lines show the profiles
computed using the global correction, while the dashed lines show the
profiles which use the local corrections. The differences in the
profiles are very minor, typically less than 0.1 mag even in our lowest
surface brightness bins.

\begin{figure}[]
\centerline{\includegraphics[width=3.0truein]{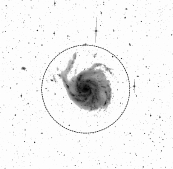}}
\caption{Our M101 B band image, showing only pixels brighter than
\mub=25 \magsec. The dotted circle shows M101's RC3 diameter of
\D25=28.8\arcmin, which significantly overestimates the isophotal size
of the galaxy.}
\label{M101R25}
\end{figure}

M101's strong asymmetry is again shown in the wide variation in surface
brightness at a given radius, typically $\sim$ 2 \magsec\ outside
R=5\arcmin. The asymmetry also leads to significant uncertainty in the
metrics that measure disk size. One canonical size metric is \R25, the
radius of the \mub=25 isophote. If we measure \R25 by calculating, on an
octant by octant basis, the outermost radius at which the mean surface
brightness falls below \mub=25, we derive values which range from
6.1\arcmin\ to 9.4\arcmin\ (Table \ref{disksize}), depending on azimuth.
These values are all significantly smaller than the isophotal size given
in the RC3 (de~Vaucouleurs \etal 1991), which lists a diameter of
\D25=28.8\arcmin. To examine this discrepancy, we re-display our B band
image in Figure \ref{M101R25}, showing only pixels brighter than
\mub=25, along with dashed circle shows the RC3 diameter. To the north
of M101, a small patch of starlight can be seen on the dashed circle at
the end of the galaxy's crooked northeastern spiral arm; if this is
strictly interpreted as the galaxy's outermost \mub=25 isophote at
R=14.4\arcmin, a doubling to convert radius to diameter would yield
the RC3 value of \D25=28.8\arcmin. Nonetheless, as a global measure of
galaxy size, Figure \ref{M101R25} clearly shows that the RC3 value
significantly overestimates M101's \mub=25 isophotal radius, by as much
as a factor of two. As an alternative measure that is less sensitive to
the morphological distortion of the galaxy, we can calculate an
areal-weighted \R25 by measuring M101's surface area at \mub$<25$ (200
square arcminutes) and converting this to an equivalent
\R25($=\sqrt{A_{25}/\pi}$) of 8.0\arcmin. We consider this a more robust
measurement of M101's isophotal size.

Similarly, the exponential scale length of the disk shows significant
variation both in azimuth and in the radial range chosen to fit. We
measure the disk scale length over two radial regions, the inner disk
(R$<$7\arcmin) and the outer disk (R$>$9\arcmin). Our choice of
R=8\arcmin\ to separate the inner and outer profiles is motivated not
only by the fact that it is roughly comparable to the \R25 values we
derive, but also because we find changes in the profile slope over the
radial range R=7\arcmin--9\arcmin\ in the various octants. Fitting the
azimuthally averaged surface brightness profile in the inner disk, we
derive a B-band scale length of 2.42\arcmin\ (Table \ref{disksize}),
similar to previous values derived from photographic imaging
(2.58\arcmin, Okamura \etal 1976). However, the scale
length of the inner disk varies strongly as a function of azimuth,
ranging from 1.9\arcmin\ to 3.4\arcmin\ across the various octants. At
large radius in the outer disk, the variation in scale length is even
greater, from 1.2\arcmin\ to 4.2\arcmin. It is interesting to note that
pure azimuthal average hides much of this variation; the fit to the
azimuthally averaged outer disk profile yields a scale length similar to
the inner disk scale length, and indeed looking at the full, azimuthally
averaged profile does not give any sense that multiple fits are
warranted. This demonstrates the inherent danger in constructing
azimuthally averaged profile; significant structural information can be
lost in the azimuthal average.

M101's color profile is shown in the lower panel of Figure
\ref{radprof}, where we plot the color profiles out to our limiting
surface brightness of \muv=29. The profiles show a strong gradient in
the inner 3\arcmin, followed by a much shallower gradient out to $\sim$
10\arcmin. The colors profiles diverge at even larger radius, with some
turning redder and some bluer. At the lowest surface brightnesses,
uncertainty in the background subtraction begins to affect the color
profiles, with color differences of up to $\sim$ 0.1 between the
profiles extracted using the ``global" versus ``local" sky estimates.
Nonetheless, the azimuthal differences in color at large radius are
larger than this uncertainty. In particular, the extremely blue colors
of the NE octant are robust against changes in the background
subtraction, and are also not systematically biased by noise in the
data. To illustrate this latter point, we have measured the color of the
background bins themselves --- which essentially measures the color
of the noise in the image --- and they come out consistently red, with
\bmv\ colors in the range 0.8--1.0. The red upturn in the outermost
radial bin in each octant therefore likely reflects the increasing
contribution of background noise to the color profiles.

\begin{deluxetable*}{cccccccc}
\tabletypesize{\scriptsize}
\tablewidth{0pt}
\tablecaption{Disk Photometric Parameters}
\tablehead{
\colhead{}       &  \colhead{}       &  \colhead{}               &\multicolumn{2}{c}{R$<$7\arcmin}           &       & \multicolumn{2}{c}{R$>$9\arcmin} \\
\cline{4-5}  \cline{7-8} \\
\colhead{Octant} & \colhead{Figure}           &\colhead{$R_{25}$} & \colhead{$h$} & \colhead{$\mu_{0,B}$} & & \colhead{$h$} & \colhead{$\mu_{0,B}$}\\
\colhead{}           & \colhead{Color}           &\colhead{[arcmin]}  & \colhead{[arcmin]}  & \colhead{[\magsec]}      &    & \colhead{[arcmin]}   & \colhead{[\magsec]} 
}
\startdata
  E &      red &  6.8 &  1.96 (0.10) & 21.13 (0.09) &  & 4.18 (0.34) & 24.25 (0.30) \\ 
SE &   orange &  6.1 &  2.08 (0.19) & 21.25 (0.14) &  & 2.48 (0.09) & 22.51 (0.20) \\ 
 S &   yellow &  7.0 &  2.71 (0.21) & 21.51 (0.09) &  & 1.38 (0.05) & 18.76 (0.30) \\ 
SW &    green &  8.5 &  3.36 (0.24) & 21.53 (0.07) &  & 1.18 (0.00) & 17.06 (0.02) \\ 
 W &     blue &  8.3 &  2.49 (0.15) & 21.36 (0.08) &  & 1.81 (0.28) & 20.60 (1.08) \\ 
NW &     cyan &  8.2 &  2.39 (0.14) & 21.47 (0.08) &  & 1.86 (0.28) & 19.92 (1.03) \\ 
 N &     gray &  6.7 &  2.13 (0.11) & 21.29 (0.08) &  & 3.84 (2.74) & 23.13 (2.59) \\ 
NE &  magenta &  9.5 &  1.92 (0.11) & 21.14 (0.10) & & 2.75 (0.31) & 21.48 (0.67) \\ 
\tableline
AZ &  black &  7.4 &  2.42 (0.05) & 21.37 (0.03) &  & 2.09 (0.06) & 21.25 (0.20)
\enddata
\tablecomments{``AZ'' gives the parameters extracted from a circular, azimuthally averaged surface brightness profile.``Figure color'' refers to the line and quadrant colors used in Figure \ref{radprof}. Uncertainties are given in parentheses.}
\label{disksize}
\end{deluxetable*}

\subsection{The NE Plume and E Spur}

\begin{figure}[]
\centerline{\includegraphics[width=3.5truein]{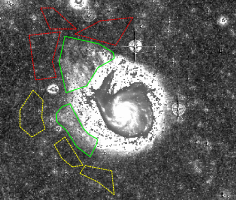}}
\caption{Regions used in photometering the NE Plume and E Spur. Regions outlined in solid
green show the NE Plume and E Spur, while sky regions for the NE Plume
and E Spur are shown in dotted red and yellow, respectively.}
\label{M101reg}
\end{figure}

To explore the outer disk of M101 in more detail, we isolate regions
defining the NE Plume and E Spur (see Figure \ref{M101reg}) and measure
the photometric properties of each region using the masked, median
binned images. As discussed in Rudick \etal (2010), at low surface
brightnesses we obtain better photometric accuracy by making a local
background subtraction, where the background is measured in regions that
are of comparable size and in close proximity to region of interest. For
each of the NE Plume and the E Spur, we identify three sky regions for
background subtraction (also shown in Figure \ref{M101reg}), and use the
variation in the regions (typically $\pm$0.2 ADU) to estimate the
uncertainty in the measurement. In photometering each region, we
restrict the measurement to pixels at low surface brightness, with $26.5
<$\mub$<29$. There are in fact almost no pixels brighter than the bright
cutoff, while the faint cutoff is chosen to reduce noise in the areal
photometry. We also apply masks around bright stars to suppress
contamination of the photometry due to diffraction spikes, column
bleeds, and other high-spatial frequency structure left behind after the
subtraction of the extended stellar wings (Slater \etal 2009). In the E
Spur, only one small mask was needed, but the NE Plume required three
large ($r=2$\arcmin) masks around very bright stars, in addition to a
handful of smaller ones around fainter stars. One of the large stellar
masks in the NE Plume covers a region of relatively high surface
brightness, meaning that our measurements of the luminosity of the
region will be slightly underestimated.

The derived photometric properties of the regions are given in Table~\ref{regprop}.
We measure a blue luminosity of L$_B$=$0.98\times10^8$ \Lsun and and
L$_B$=$0.78\times10^8$ \Lsun for the NE Plume and E Spur, respectively. As
suggested by the color map shown in Figure \ref{M101multi} and the color
profiles in Figure \ref{radprof}, the \bmv\ color of the NE Plume is
quite blue, \bmv=0.21, while the E Spur is redder, \bmv=0.44. The color
measured from the derived B and V luminosities is, of course, luminosity
weighted, and could be biased towards the highest surface brightness
regions. To guard against such a bias, we also calculate an areal
weighted color, defined as the median color of all pixels in the region,
and find no significant difference between the areal weighted and
luminosity weighted colors.

The very blue \bmv\ color of the NE Plume is striking compared to the
integrated colors of galaxies. Colors this blue are typically found only
in irregular galaxies, and even in those systems they represent the blue
extreme (Fukugita \etal 1995; Hunter \& Elmegreen 2006). Even gas-rich
low surface brightness galaxies, typically thought to be relatively
unevolved systems, show \bmv\ colors which are typically much redder,
with \bmv$>$ 0.4 (Bothun \etal 1997). Given its extremely
blue color, it is hard to envision the NE Plume hosting any significant
population of old stars, and instead must be dominated by younger
populations. We will quantify this argument and explore various star
formation histories for the NE Plume using population synthesis models
later in this section.

\begin{deluxetable}{ccc}
\tabletypesize{\scriptsize}
\tablewidth{0pt}
\tablecaption{Region Properties}
\tablehead{\colhead{ } & \colhead{NE Plume} & \colhead{E Spur}}
\startdata
Area [$10^5$ arcsec$^2$] & 5.67 & 2.74 \\
Area [kpc$^2$] & 695 & 336 \\
L$_B$ [10$^8$ L$_{\sun,B}$] & 0.98 $\pm$ 0.003 & 0.78 $\pm$ 0.01  \\
L$_V$ [10$^8$ L$_{\sun,V}$] & 0.66 $\pm$ 0.03 & 0.65$\pm$ 0.01  \\
B-V (lum wt) & 0.21 $\pm$ 0.05 & 0.44  $\pm$ 0.02\\
B-V (areal wt) & 0.20 $\pm$ 0.05 & 0.45 $\pm$ 0.02 \\
M$_{\rm HI}$ [$10^8 M_\sun$] & 6.4 (8.1) & 2.0 (2.0)  \\
$\Sigma_{\rm HI}$ [$M_\sun / pc^2$] & 1.9 (2.0) & 1.1 (1.1) \\
m(FUV) & 15.64 (15.14) & 16.62 (16.58) \\
SFR$_{\rm FUV}$ [M$_\sun$/yr] & 0.015 (0.023) & 0.006 (0.006) 
\enddata
\tablecomments{In all cases, properties are measured for the regions after applying
the star masks corresponding to the optical data. In the case of the THINGS \hi data
and GALEX far ultraviolet data, properties measured without the optical star masks
are given in parentheses.}
\label{regprop}
\end{deluxetable}

For a more complete view of the physical properties of these regions, we
also measure the \hi\ mass and FUV flux from the THINGS and GALEX
datasets, respectively. In doing so, the question arises whether or not
to apply the same stellar masks to the multiwavelength data that we used
for the optical data. Of course, the contamination to the data from
these stars is minimal in the GALEX data, and non-existent in the \hi\
data. However, applying the masks treats all the datasets the same and
ensures the most direct comparison between the extracted photometric
parameters. To balance these considerations, in the GALEX and THINGS
datasets, we measure the properties of each regions both with and
without applying the masks, and report both sets of numbers in Table
\ref{regprop}. 

For each region, we measure the total FUV flux from the GALEX image,
subtracting off a background level obtained using the same background
regions that were used in the optical photometry. We obtain total FUV
magnitudes of $m_{AB}=15.64\ (15.14)$ and $m_{AB}=16.62\ (16.58)$ for
the NE Plume and E Spur, respectively, with the numbers in parentheses
reflecting measurements without the optical star masks being applied. As
the FUV emission arises from young stellar populations with typical ages
100 Myr or less, it traces recent star formation in each region. We can
convert these fluxes to integrated star formation rates using the
prescription of Hunter \etal (2010), which corrects
Kennicutt's (1998) relationship between star formation rate and
ultraviolet flux for use in low metallicity environments (as might be
expected for the outer regions of M101; Kennicutt \etal
2003). This conversion yields very low star formation rates of 0.015
(0.023) \sfr and 0.006 (0.006) \sfr for the NE Plume and E Spur,
respectively. These star formation rates are approximately 30 times
lower than the upper limits set by the lack of \ha flux in these regions
from the Hoopes \etal (2001) dataset (Figure \ref{M101multi}). 

Using the THINGS \hi\ map, we measure total \hi\ masses of
$6.4\times10^8$ \Msun and $2.0\times10^8$ \Msun for the NE Plume and E
Spur, respectively, and again these are likely underestimates of the
total \hi\ mass since both regions extend slightly beyond the extend of
the THINGS map. The corresponding mean surface density of the gas is 1.9
and 1.1 \Msun~pc$^{-2}$ in each region. Using the kinematic threshold
density argument of Kennicutt (1989), we can calculate the expected
critical density for star formation as
$$\Sigma_{crit} = 0.59\alpha V_{c} / R $$
where $V_{c}$ is the circular velocity in km s$^{-1}$, $R$ is the radius
in kpc, $\alpha$ is a unitless parameter $\sim$ 0.67, and where we have
used the flat rotation curve approximation for the relationship. Using a
circular velocity of 220 km s$^{-1}$ (Bosma \etal 1981) and a
characteristic radius of $\sim$ 32 kpc for each region, we calculate a
critical density for star formation of 2.7 \Msun~pc$^{-2}$. Both regions
fall below this threshold density, explaining the lack of significant
star formation in each region. The fact that the NE Plume has higher
\hi\ surface density, closer to the critical density, may have made it
more responsive to any dynamical perturbation, such as a tidal
interaction, and result in the somewhat higher star formation rate
compared to that seen in the E Spur.

We can also use the optical properties of the regions to make a simple
estimate of the time-averaged star formation rates needed to produce the
optical luminosity we see. Using GALEV population synthesis models
(Kotulla \etal 2009), we first construct constant star formation stellar
populations, employing a Kroupa (2001) IMF and fixed sub-solar
metallicities of [Fe/H]=--0.3 and --0.7, similar to what is expected for
the outer disk of M101 (Kennicutt \etal 2003). We determine the time at
which the model color matches the observed color of each region, and use
the model mass-to-light ratio at that best-match time to convert the
observed stellar luminosity to a stellar mass. The inferred star
formation rate is then calculated as the stellar mass divided by
the population age. For the NE plume, the population age and stellar
mass-to-light ratio inferred from the color are $t=$1.5 Gyr and
M/L$_V$=2.0, implying a star formation rate of 0.09 \sfr\llap, six times
greater than that inferred by the FUV luminosity. In other words,
a constant star formation model, when matched to the color of the
NE Plume, significantly under-predicts the amount of stellar light. 
A longer duration of star formation would eventually build up enough
stellar luminosity, but at a very late time where the population would
be significantly redder than observed.

Given that a constant star formation model cannot explain the photometric
properties of the NE Plume, we turn to models involving a burst of star
formation, as might be expected if star formation had been triggered
by a recent accretion or interaction event. We use GALEV models as before,
this time with star formation histories modeled as a Gaussian bursts with
peak star formation rates $I_0$ and durations characterized by a Gaussian
width $\sigma$. We explore a range of burst durations, ranging $\sigma$=25 
Myr to $\sigma$=1 Gyr. We compare the color and luminosity of each model
as a function of time to the observed color and luminosity of the NE Plume,
dynamically setting $I_0$ to the proper value to yield the UV measured
star formation rate at the present time. While the constraining power of the
data is limited given the additional free parameters of these models, we
find that moderate starbursts with peak intensities of $I_0 \sim 0.1-0.2$ \sfr
and durations of $\sigma \sim 75-100$ Myr, observed $\sim 250-350$ Myr
past peak do a reasonable job of simultaneously matching both the color 
and luminosity of the NE Plume.  

Performing a similar analysis for the E Spur, we find that its much
redder color means that a constant star formation history cannot achieve
the observed color until an unrealistically late time of 13--15 Gyr. The
star formation rate implied by the optical color and luminosity is 0.05
\sfr, an order of magnitude greater than the FUV-derived star formation
rate. Here too, clearly a constant star formation model will not work;
even over a Hubble time, the current star formation rate cannot build up
enough luminosity to match that of the E Spur. Unfortunately, its red
color makes the E Spur less amenable to any more complicated
analysis. Many different star formation histories can produce redder
colors, such as old bursts and declining or truncated star formation
rates, particularly when combined with the possible effects of dust. A
number of plausible scenarios exist --- the E Spur could be material
pulled out from the inner disk due to a tidal interaction, or it could
simply be the continuation of a spiral arm into the outer disk.
Unfortunately, our broadband imaging data simply do not have the
constraining power to discriminate between these various possibilities.

We have confined our analysis to the outer, tidally disturbed regions of
M101 where the dust content is low (\eg Popescu \etal 2005) and, at
least in the case of the NE Plume, the broadband colors are quite blue.
Extending the analysis to the inner, redder high surface brightness
regions of the galaxy would be significantly complicated by dust
extinction and reddening, and yield ambiguous results. However, it is
interesting in this context to compare our results to the study of
Bianchi \etal (2005) who used GALEX UV colors combined with SDSS optical
data to study the recent star formation history of M101's inner disk
($r<$10\arcmin). They find UV colors that inconsistent with a constant
star formation model, and instead argue for young populations a few
hundred million years in age. While these constraints apply to the inner
disk, they are qualitatively similar to what we infer from our analysis:
young stellar populations in the disk outskirts born in a recent star
formation event.

\section{Discussion}

Our deep optical imaging of M101 has revealed the structure of its outer
disk, tracing starlight out to a radius of nearly 50 kpc. The well-known
asymmetry in the structure of the high surface brightness, inner parts
of the galaxy strengthens at large radius and lower surface brightness,
but in a position angle that slews 180\degr\ from that of the inner
disk. In the outer disk we have identified two low surface brightness
features, the NE Plume and E Spur. While the redder colors of the E Spur
make it hard to place strong constraints on the stellar populations and
dynamical history, the much bluer NE Plume must have formed from very
recent star formation in M101's outer disk. The presence of a
significant amount of blue optical light in the NE Plume extends the
inferred lifetime for outer disk star formation from the $\sim$ 100 Myr
scales traced by the GALEX UV emission (Thilker \etal 2007) to
timescales of $\sim$ Gyr.

The ubiquity of star formation in the outskirts of disk galaxies has
been shown in a variety of studies, using imaging in \ha (\eg Ferguson
\etal 1998; van Zee \etal 1998; Goddard \etal 2010) and far ultraviolet
(Gil de Paz 2005; Thilker \etal 2005, 2007). Even our own Galaxy shows
evidence for such star formation in the outer disk (de Geus \etal 1993; 
Kobayashi \& Tokunaga 2000; Carraro \etal 2010). M101 has been
classified a Type I extended ultraviolet (XUV) disk galaxy (Thilker
\etal 2007), where structured UV emission is seen well outside the
nominal optical size of the disk. A variety of mechanisms have been
proposed for exciting such activity, including cold accretion of gas
from the surrounding environment, the effects of spiral waves
propagating from the inner disk and driving instabilities in the disk
outskirts, the accretion of a gas-rich companion, and
interaction driven perturbations of existing gas at large radius.

Under cold inflow models, gas condensing from the intergalactic medium
can accrete directly onto the outer disk (\eg Kere\v s \etal 2005, 2009;
Dekel \& Birnboim 2006), triggering extended star formation in the disk
outskirts. This accretion may be quite lopsided (Stewart \etal 2011),
and drive disk asymmetries (Bournaud \etal 2005) similar to those we see
in M101. However, while M101 (as well as many other galaxies; Sancisi
\etal 2008 and references therein) shows extended HI signatures reminiscent
of those seen in cold inflow models, whether this accretion involves
truly fresh, primordial gas or gas ejected from the host or companion
galaxy during an interaction remains unclear. Furthermore, while the
morphological similarities with the cold accretion model are intriguing,
cold mode accretion is predicted to shut down for massive (M$_{tot} >
10^{12}$ \Msun) galaxies at low redshift, after they have build up a hot
gaseous halo that impedes the flow of cold gas to the disk (Dekel \&
Birnboim 2006). While M101 does not host a massive hot halo (Kuntz \etal
2003; Kuntz \& Snowden 2010), even a fairly tenuous halo of hot gas
could be capable of limiting direct cold inflow. It is unlikely,
therefore, that a massive galaxy like M101 would be experiencing any
significant cold flow accretion at the present day.

The spiral-driven disk instability model also seems not to explain the
structure we see at large radius in M101. Under this model (Bush \etal 2010), star
formation in the outer disk should be an ongoing (albeit sporadic) event
over the lifetime of the disk, yielding a component of extended
starlight that is distributed more evenly in azimuth. This is contrary to
what we see in M101, where the extended light is not at all
axisymmetric, nor does it have colors indicative of an extended star
formation history. Furthermore, in the spiral-driven instability model,
the outer disk star formation should trace the spiral pattern of the
galaxy (Bush \etal 2010); while this may be the case for the E Spur, the
NE Plume extends outwards perpendicular from M101's strong NE spiral
arm, rather than tracing it to the north. The asymmetry of the outer
disk and the morphology of the NE Plume makes a poor match for the
spiral-driven instability model.

In contrast, the external accretion model for XUV disks fares somewhat
better, in that accretion of cold gas would be expected to be more
sporadic and asymmetric; a recent accretion event could lead to the
structure we see in M101. Deep 21-cm imaging of M101 (\eg Huchtmeier \&
Witzel 1979; Allen \& Goss 1979; van der Hulst \& Sancisi 1988; Walter
\etal 2008; Mihos \etal 2012) has shown that the HI distribution around
M101 is quite extended and asymmetric, including a very extended plume
of material to the southwest at very low column density. Once pulled
out, tidal debris typically falls back to the host galaxy over an
extended period of time (\eg Hibbard \& Mihos 1995), feeding the outer
disk and perhaps triggering star formation as it falls back in. Indeed,
the two streams of high velocity gas in M101 first noted by van der
Hulst \& Sancisi 1988 are spatially coincident with the NE Plume and E
Spur (Kamphuis 2008; Sancisi \etal 2008), arguing that these features
may have been triggered by recent gas infall from outside M101's disk.

Even if not driven by direct accretion onto M101's disk, the extended
star formation seen in the NE Plume could be triggered by the tidal
effects of an interaction on pre-existing gas in the outer disk. At the
HI surface densities found in in this region (1--2 \Msun pc$^{-2}$),
star formation is normally greatly suppressed (Kennicutt 1989; Bigiel
\etal 2008), likely due to the difficulty of making molecular gas at low
density (Krumholz \etal 2009). While the current star
formation activity (as traced by UV and \ha) is quite low, it must have
been higher in the past to account for the amount of blue light seen in
our deep imaging, meaning that the molecular conversion efficiency must
have been higher as well. The caustics that form during galaxy
encounters (\eg Struck \etal 2011) are an effective way at
compressing gas and driving the formation of molecular clouds, and are
also often sites of enhanced star formation in obviously interacting
galaxies (Keel \& Wehrle 1993; Kaufman \etal 1999; Elmegreen \etal
2000); what we see in M101 may simply be a milder or older version of
those more dramatic instances. That the morphology of the NE Plume and E
Spur are similar to those seen in simulations of flyby interactions (\eg
Toomre \& Toomre 1972; Howard \etal 1993) also argues for an
interaction-induced origin. In this scenario, while the interaction
may be associated with additional accretion, the extended star
formation seen in the NE Plume is induced in the gas pre-existing in
M101's disk, rather than from gas being accreted externally.

If the young features we see in the outer disk are in fact
interaction-driven, the clear question is: who is doing the driving? One
obvious culprit is the dwarf companion NGC 5477, which currently sits
22\arcmin\ (44 kpc) east of M101, between the NE Plume and E Spur.
Indeed, the morphology of these features is similar to that expected
when a satellite galaxy on a prograde orbit opens a gap in the galactic
disk (\eg Hernquist \& Mihos 1995; Walker \etal 1995).
This could explain the structure of the low surface brightness outer
disk, and also perhaps the high velocity gas seen associated with the
features, as a small companion embedded in the outer disk could easily
excite non-planar motions. However, it would seem difficult for a low
luminosity dwarf like NGC 5477 to drive such a strong response in the
{\it inner} regions of a massive Sc spiral like M101 as well. The
luminosity ratio between NGC 5474 and M101 is $\sim$ 250:1 (de Vaucouleurs
\etal 1991); if this tracks the mass ratio of the pair, this is an even
more low mass interaction than that between the Milky Way and the LMC,
and yet the damage done to M101 is significantly more pronounced than
anything the LMC has imprinted on the Milky Way.

Another possible interaction partner is NGC 5474, projected 44\arcmin\
(88 kpc) to the south, but the evidence here is mixed as well. NGC 5474
is more luminous than NGC 5477, making it likely more able to drive a
stronger response in M101, but the luminosity ratio between it and M101
(17:1 in V-band) would still classify it as a low mass
encounter.\footnote{Because both M101 and NGC 5477 are largely face-on
and, in the case of M101, kinematically disturbed, deriving dynamical
masses for the two galaxies is quite difficult. See, for example, Bosma
\etal 1982 and Rownd \etal 1994.} NGC 5474 does sport an off-center
central bulge, but is otherwise morphologically undisturbed, even its
very low surface brightness outer regions (Figure \ref{Bmosaic}).
Similarly, its kinematic structure shows no strong deviations from pure
circular rotation (Rownd \etal 1994), despite the off-center bulge.
However, deep 21-cm mapping traces gas between the two galaxies at
intermediate velocities (Mihos \etal 2012), as might be expected from a
close passage between the two galaxies. If we take the timescale for the
starburst in the NE Plume (250--350 Myr ago) as the time of closest
approach, and if the encounter is slow ($\sim V_c$), as needed to drive
a strong response in M101, the timescales would place NGC 5474 $\sim$
100 kpc from M101 today, quite similar to its observed (projected)
distance.

Ultimately, however, the story to M101's dynamical history may be more
complex than a single interaction scenario. It may be that both a flyby
interaction with a moderately massive galaxy (NGC 5474) and continued
dynamical forcing from a satellite companion (NGC 5477) may be needed to
describe M101's full morphological and kinematic structure. Previous
simulations by Combes (1991) showed that the high velocity gas in M101
could be explained by an interaction with a companion galaxy, but did
not attempt to explain M101's overall asymmetry, any large scale tidal features,
or the properties of its companions. With a wealth of new
constraining kinematic and photometric data existing over a wide range
of wavelength, new dynamical modeling of the M101 system is clearly
warranted.

It is also interesting to ask how the many dynamical processes we see at
work in M101 may be reshaping its disk structure over the long term,
particularly since much attention has been paid to using differences
between the inner and outer scale lengths of spiral galaxies as a tracer
of disk evolution. Disk profiles that steepen in the outskirts are
referred to as ``down-bending'' profiles, while surface brightness
profiles which flatten with radius are designated ``up-bending" (van der
Kruit 1979, Pohlen \etal 2002, 2007). Growing evidence suggests that
structure in the radial profile of disk galaxies correlates with Hubble
type: up-bending profiles are more commonly found in early type disks,
while late type spirals (like M101) are more likely to posses
down-bending profiles (Pohlen \& Trujillo 2006; Erwin \etal
2008; Gutierrez \etal 2011). The cause for this diversity in
profiles is quite unclear; possible mechanisms include radial
truncations in star formation (Kennicutt \etal 1989), dynamical
responses to minor mergers (Younger \etal 2007), and radial migration of
stars (Sellwood \& Binney 2002; Debattista \etal 2006; Roskar \etal
2008ab). Many of these processes are undoubtedly at work in M101. The
induced star formation at large radius see in the NE Plume can build the
outer disk stellar populations; interactions with companions can warp
and spread existing disk populations, perhaps explaining the E Spur; and
M101's strong asymmetry and spiral structure are sure to drive
scattering and outward radial migration of stars from the inner disk.

An examination of M101's radial profiles (Figure~\ref{radprof}) shows
that, as a function of azimuth, M101 currently possesses {\it both}
types of profiles -- down-bending to the southwest and up-bending to the
northeast. However, while now very asymmetric, as the galaxy evolves
much of this light will mix radially and azimuthally, smoothing the disk
profile. In this context, it is interesting that our {\it azimuthally
averaged} disk profiles show no obvious break in either surface
brightness or color. Azimuthally averaged, the separate inner and outer
profile fits suggest a very slight down-bending profile, but the
motivation for separate fits is weak -- a single exponential could just
as easily fit the profile. In M101, the effects of the extended star
formation and the tidal perturbation appear to be building the outer
disk by adding or moving stars to large radius. If at earlier times,
M101 had a down-bending profile, characteristic of late type spirals
(Pohlen \& Trujillo 2006) and as seen in the southwest side of the disk,
that profile could be being erased by these processes. With M101 living
in a group environment, future interactions are likely, and may continue
driving a slow evolution of M101's disk profile from down-bending to
up-bending. If repeated interactions also drive a morphological
transformation towards earlier Hubble type, this would explain the
connection between morphological type and disk profile shape.
Particularly since these types of interactions should be common for
galaxies in the group environment, we are likely catching M101 in an
important phase of galaxy evolution as it continues building its outer
disk through triggered star formation and the effects of interactions
with its companions.

Of course extrapolating the current profile shape into the future
depends not only on effective dynamical mixing, but also on the radial
distribution of the underlying stellar population that produces the
light. If, for example, the outskirts are dominated by young stars while
the inner regions have a mix of stellar ages (and ongoing star
formation), the disk may well end up with a down-bending profile as the
outer population reddens and fades more rapidly than the inner
population. Unfortunately, with only broadband imaging in two bands, our
constraints on the stellar populations in M101's outer disk are limited.

One technique that could yield more information about the stars in
M101's outer disk is direct imaging of the discrete stellar populations
in these regions. Such studies using the {\it Hubble Space Telescope} in
other nearby galaxies have revealed a wealth of information about the
ages and metallicities of the stellar populations (Dalcanton \etal 2009,
2012; Williams \etal 2011; Radburn-Smith \etal 2011, 2012). The tip of
the main sequence and the presence of ``blue loop" stars transiting from
the main sequence to the red giant branch can place strong constraints
on the stellar population ages for populations younger than about 300
Myr, while for populations older than about 5 Gyr, the shape of the red
giant branch constrains the metallicity. Better ages would help refine
the timescale and duration of the starburst in the NE Plume, while the
metallicity could place constraints on whether these populations were
drawn from the inner, more metal-rich disk environment or, instead,
formed {\it in-situ} from low metallicity gas. At the distance of M101,
the tip of the main sequence would be at $m_I \sim$ 27.0--28.5 for
population ages 200--300 Myr old, while for older populations the tip of
the red giant branch is at $m_I \sim 26.0$, making it a feasible study
using deep HST or JWST imaging.

\section{Summary}

We have used deep ($\mu_{B,lim}=29.5$), wide field (6 square degrees)
imaging of M101 and its surrounding environment to study the structure
of and stellar populations in M101's outer disk and to search for
interaction signatures in M101 and its companion galaxies. We trace
starlight in M101's disk out to nearly 50 kpc, three times the \R25
optical radius of the galaxy. The strong $m=1$ asymmetry in M101's disk
slews 180\degr in azimuth as a function of radius, with the inner disk
being distorted to the southwest and the outer disk to the northeast. We
identify two structures in M101's outer disk, the very blue (\bmv=0.21)
NE Plume and the redder (\bmv=0.44) E Spur. Aside from these features,
we see no extended optical tidal debris around M101 or its companions as
might be expected to result from a close passage of a massive companion.
Nonetheless, the marked asymmetry of both M101 and NGC 5474, and the HI
tidal signatures seen in M101 at larger radius (Huchtmeier \& Witzel
1979; van der Hulst \& Sancisi 1988; Mihos \etal 2012) all argue for
ongoing interactions with lower mass companions in the environment of
the group.

We construct radial surface brightness and color profiles for M101's
disk, both azimuthally averaged and, due to M101's strong asymmetry, as
a function of azimuth. While the azimuthally averaged surface brightness
profile shows a smooth exponential (with scale length $h\sim2.2$\arcmin)
out to R=15\arcmin, azimuthal averaging hides much of the interesting
structure in M101's disk. As a function of azimuth, M101's surface
brightness profile exhibits both truncated and anti-truncated structure.
The disk shows a steep color gradient in the inner 3\arcmin, followed by
a much shallower gradient out to R$\sim 10$\arcmin. Outside this radius
the different color profiles diverge, with some turning red (\bmv $\sim$
0.5) and others staying quite blue (\bmv $\sim$ 0.2--0.3). The strong
asymmetry of the disk makes a unique determination of the \R25 optical
radius impossible, with values ranging from \R25=6.1\arcmin--9.4\arcmin;
we recommend using an areal-weighted value of \R25=8.0\arcmin\ over the
RC3 value of \R25=14.4\arcmin.

As the dominant member of a small group of galaxies, M101 gives us a
view of processes shaping the evolution of disk galaxies in the group
environment. The disk's strong asymmetry and azimuthally varying color
and luminosity profiles show a galaxy building up the stellar
populations in its outer disk, likely driven by fly-by interactions with
its less-massive companion galaxies. These interactions can build the
outer disk by a variety of mechanisms. Direct accretion of material from
the companion can deposit stars and gas into the disk outskirts, and the
spatial coincidence between the extended NE Plume and E Spur and the
high velocity HI clouds in M101's disk (van der Hulst \& Sancisi 1988)
may be a signal of such activity. However, even in the absence of direct
accretion, interactions can grow the outer disk by triggering new star
formation in the galaxy's extended HI disk, and indeed the blue colors
of the NE Plume argue for a burst of star formation $\sim$ 250--350 Myr
ago, possibly dating a recent encounter with NGC 5474, 44\arcmin\ (88
kpc) to the south of M101. The redder E Spur may also have formed during
this interaction, from material drawn out from the inner disk of M101.
Alternatively, or perhaps concurrently, an interaction with nearby
(22\arcmin, 44 kpc) dwarf companion NGC 5477 may also drive some of the
response seen M101's outer disk. These encounters all act to dynamically
heat the disk and move stars from the inner regions to the disk
outskirts, either on rapid timescales as tidal features form during the
initial encounter, or on longer timescales as the self-gravitating
response of the disk drives radial migration of stars outwards (\eg
Sellwood \& Binney 2002; Debattista \etal 2006; Roskar \etal 2008ab). In
M101, we are likely witnessing the growth and reshaping of its outer disk 
due to these processes, triggered by interactions in the group environment.

\acknowledgments

This work has been supported by the NSF through grants AST-0607526 and
AST-1108964 to J.C.M. and AST-0807873 to J.J.F. We thank Stephanie Bush, Kelly
Holley-Bockelmann, Pat Durrell, and Stacy McGaugh for several helpful
discussions. This research has made use of the NASA/IPAC Extragalactic
Database (NED) which is operated by the Jet Propulsion Laboratory,
California Institute of Technology, under contract with the National
Aeronautics and Space Administration.

{\it Facility:} \facility{CWRU:Schmidt}

\end{document}